\documentclass[aps,prd,twocolumn,groupedaddress,nofootinbib]{revtex4}

\usepackage{revsymb,epsfig,verbatim}
\usepackage{ulem}

\def\plotsdir{\string }  

\def\be {\begin{equation}}
\def\ee {\end{equation}}
\def\bea {\begin{eqnarray}}
\def\eea {\end{eqnarray}}
\def\bean {\begin{eqnarray*}}
\def\eean {\end{eqnarray*}}

\newcommand{\bfr}{{\bf r}}
\newcommand{\bfs}{{\bf s}}

\newcommand{\bfd}{{\bf d}}
\newcommand{\bfl}{{\mbox{\boldmath $\ell$}}}

\def\gsim{ \lower .75ex \hbox{$\sim$} \llap{\raise .27ex \hbox{$>$}} }
\def\lsim{ \lower .75ex \hbox{$\sim$} \llap{\raise .27ex \hbox{$<$}} }
\def\\{\hfil\break}
\def\spose#1{\hbox to 0pt{#1\hss}}
\def\lta{\mathrel{\spose{\lower 3pt\hbox{$\mathchar"218$}}
     \raise 2.0pt\hbox{$\mathchar"13C$}}}
\def\gta{\mathrel{\spose{\lower 3pt\hbox{$\mathchar"218$}}
     \raise 2.0pt\hbox{$\mathchar"13E$}}}

\newcommand{\Mvir}{M_{\rm vir}}
\newcommand{\Rvir}{r_{\rm vir}}
\newcommand{\Tvir}{T_{\rm vir}}
\newcommand{\ypoly}{y_{\rm poly}}

\newcommand{\geff}{\gamma_{\rm eff}}
\newcommand{\Mnl}{M_{\ast 0}}

\begin{document}



\title{Detection of Hot Gas in Galaxy Groups via the Sunyaev-Zel'dovich Effect}


\author{K.~Moodley$^{1}$, R.~Warne$^{1}$, N.~Goheer$^{2}$, H.~Trac$^{3}$}
\affiliation{$^{1}$Astrophysics and Cosmology Research Unit, School of Mathematical Sciences, University of KwaZulu-Natal, Durban, 4041, South Africa\\
  $^{2}$ Department of Mathematics and Applied Mathematics, University
  of Cape Town,
  Rondebosch, 7700, South Africa\\
  $^{3}$ Harvard-Smithsonian Center for Astrophysics, Cambridge, MA
  02138, USA}


\date{\today}

\begin{abstract}

Motivated by the observed shortfall of baryons in the local
universe, we investigate the ability of high resolution cosmic microwave
background (CMB) experiments to detect hot gas in the outer regions of
nearby group halos. We construct hot gas models with the gas in hydrostatic 
equilibrium with the dark matter and described by a polytropic equation of state.
We also consider models that add entropy to the gas in line with constraints from
X-ray observations. We calculate the thermal Sunyaev-Zel'dovich (SZ) signal in these 
halos and compare it to the anticipated sensitivities of forthcoming SZ survey experiments
such as ACT, PLANCK and SPT. Using a multi-frequency Wiener filter we derive SZ 
detectability limits as a function of halo mass and redshift in the presence of  
galactic and extragalactic foregrounds and the CMB. We find that group-sized halos
with virial masses below $10^{14} M_\odot$ can be detected at $z ~\lsim~ 0.05$
with the threshold mass dropping to $3-4\times 10^{13} M_\odot$ at $z~\lsim~0.01$.
The SZ distortion of nearby group-sized halos can thus be mapped out to the virial radius 
by these CMB experiments, beyond the sensitivity limits of X-ray observations. 
These measurements will provide a unique probe of hot gas in the outer regions of group 
halos, shedding insight into the local census of baryons and the injection of entropy
into the intragroup medium from non-gravitational feedback.

\end{abstract}

\pacs{}

\maketitle

\section{Introduction}
\label{section:intro}

Through the distinctive signatures that cosmological parameters, such
as the baryon density, matter density and spatial curvature have on
the cosmic microwave background (CMB) anisotropy, recent measurements
of the CMB temperature and polarization spectra \citep[][and references therein]{nolta_2008}
have placed precise constraints on these parameters, thereby
establishing a cosmological model that is consistent with a wide range
of astronomical observations \citep[][and references therein]{dunkley_2008}. 
Moreover, because the physics underlying the
CMB anisotropy is to a good approximation linear on these scales, the
CMB provides a powerful probe into the physics of the early universe
and the primordial perturbations \citep[][and references
therein]{komatsu_2008}.

Whereas on large angular scales the CMB anisotropy is fairly well
described, apart from a late-time integrated Sachs-Wolfe (ISW)
contribution \cite{sachs_wolfe_1967}, by the geometrical projection of
inhomogeneities at the last scattering surface onto our celestial
sphere, the so-called primary anisotropies, on smaller angular scales
of around an arcminute CMB photons are more significantly affected by
gravitational and scattering processes along the line of sight. These
secondary anisotropies provide an effective probe of the physics of
the low-redshift universe. 

The most significant secondary anisotropy on scales of around a few
arcminutes comes from inverse Compton scattering of CMB photons off
hot gas in galaxy clusters along the line of sight, the so-called
thermal Sunyaev-Zel'dovich (SZ) effect \cite{sunyaev_zeldovich_1970}.
The population of hot electrons in the cluster gas
imparts energy to photons in the Rayleigh-Jeans part of the spectrum
pushing them into the Wien tail. This distorts the thermal CMB
spectrum by creating a deficit of photons at low frequencies and an
excess of photons at high frequencies with no distortion at the SZ
null frequency, $\nu \simeq 218$ GHz. The galaxy cluster appears as a cold spot on the sky
at frequencies below the SZ null and as a hot spot above the SZ
null. The unique spectral signature of the SZ effect will allow
multi-frequency, high-resolution CMB experiments to make a relatively
clean separation of the thermal SZ effect from the primary CMB and other
contaminants such as point sources and galactic foregrounds.

The utility of the thermal SZ effect as a cosmological
probe arises from the fact that the Compton distortion of the CMB is a
scattering effect, so that the central decrement towards a cluster is
independent of the redshift of the cluster. Moreover the angular
diameter distance flattens out at $z\sim 1$ so that
the angular size of the cluster is approximately constant at high
redshift. This means that the SZ effect does not suffer from the
strong redshift dimming of its optical and X-ray counterparts with the
consequence that a microwave background SZ cluster survey can detect 
a larger proportion of lower mass
clusters out to higher redshifts. The thermal SZ effect can thus be used to
construct galaxy cluster catalogs with a well defined selection function,
thereby providing a potentially powerful probe of cosmology through 
the evolution of the cluster abundance (for a review see \cite{carlstrom_2002}).

While most work has focussed on the detection of the SZ effect in
galaxy clusters and its cosmological application, there is an
intriguing possibility that high resolution CMB experiments with
sufficient sensitivity may be able to detect hot gas in lower mass
galaxy or galaxy group halos. Whereas the central SZ distortion scales
roughly in proportion to the mass of the halo, the total SZ flux, $S_{\rm SZ}
\propto M^{5/3}/d_A^2,$ has an additional dependence on the angular
diameter distance of the halo. This means that nearby low mass halos with a large
angular size could produce a significant SZ flux. This fact was
exploited in \cite{taylor_2003} to study the possibility of detecting
the SZ effect in the halo of our neighbouring galaxy, M31, with the
upcoming PLANCK surveyor. In this paper we focus on how well current
and forthcoming CMB experiments will detect the SZ effect in nearby
galaxy and group halos. Such a detection will provide unique insights
into the distribution of baryons and the properties of hot gas in the
outskirts of galaxy and group halos. 

Indeed, in a census of baryons in the local universe
\cite{fukugita_1998, fukugita_peebles_2004} it has been argued that
most of the uncertainty in the baryon budget comes from the
uncertainty in the mass in ionized plasma associated with groups of
galaxies.  Whereas the baryon fraction inferred from light element
abundances (e.g., \cite{steigman_2007}), the CMB anisotropy \cite{dunkley_2008} 
and the high redshift Ly$\alpha$ forest \cite{kirkman_2003} give a consistent baryon fraction of
$\Omega_b \approx 0.044,$ only about ten percent of these baryons are
observed to be in stars, hot gas in clusters, and neutral and
molecular gas in galaxies at low redshift. Cool plasma in and around
groups of galaxies indirectly detected through quasar absorption lines
\cite{penton_2000, penton_2004} makes up twenty to twenty five percent
of the baryon budget, while the warm-hot gas residing in the
filamentary large-scale structure is believed to account for another
thirty to thirty five percent of the baryons \cite{dave_2001,
cen_ostriker_2006}. The remaining thirty to forty
percent of baryons is likely to be tied up in galaxy groups in a low
density, warm-hot plasma.  Recent X-ray observations \cite{Sun_2008}
that report average gas fractions, $f_{\rm gas} \approx 0.08,$ out to
$r_{500}$ in galaxy groups may have detected roughly half of these baryons.
When combined with the fact that the gas fraction profiles inferred
from these measurements are still rising at $r_{500}$ (see also
\cite{Vikhlinin_2006}) this indicates that a large fraction, perhaps
as much as fifteen to twenty percent, of baryons in the universe may
be in the outskirts of galaxy halos and in the intragroup medium
beyond $r_{500}$. Recent evidence for this warm-hot component has come
from quasar absorption lines in the ultraviolet (O VI; e.g.,
\cite{tripp_2000, sembach_2003, tumlinson_2005}) and in X-rays (OVII
and OVIII; e.g., \cite{fang_2003}) observed around galaxies and groups
of galaxies but these detections only map the two-dimensional gas
distribution along a few lines of sight to allow a complete study of
its spatial properties.

More detailed observations of this warm-hot component will yield important clues 
into the impact of galactic feedback on the distribution of gas in the intragroup 
medium. The relative dearth of hot gas observed in the central regions of galaxies 
and galaxy groups results in these halos being less X-ray luminous for a
given temperature as compared to more massive galaxy clusters. The
resulting break in the luminosity-temperature relation observed in
X-ray clusters and groups \cite{McCarthy_2004} suggests that
non-gravitational processes that modify the entropy structure in the
central regions of low mass clusters and groups are responsible for
departures from the cluster scaling relations expected in self-similar
models. Several models have been proposed to explain the excess
entropy (for a recent review see \cite{Voit_2005}), including
pre-heating of infalling gas before it was shock-heated to the virial
temperature, radiative cooling of low entropy gas to form stars, and
feedback from supernovae and active galactic nuclei that increased the
temperature and reduced the density of gas in the central regions of
galaxies and groups.  While a combination of radiative cooling and
feedback seems to reproduce the central entropy excess in low mass
clusters and groups, it is clear that improved measurements of the
properties of hot gas in these halos are necessary to understand the
exact details of entropy injection. Whereas X-ray observations lack
sensitivity to the hot gas beyond $r_{500}$ because the X-ray
luminosity scales as the square of the (decreasing) gas density, the
SZ effect scales linearly with the density and thus provides a
potentially more sensitive probe of the hot gas in the outskirts of
galaxy and group halos. Moreover, the SZ effect measures the projected
gas pressure which is a complementary probe to the X-ray luminosity
and can provide new constraints on models of the intracluster and
intragroup medium.  While the SZ effect has been detected in a number
of galaxy clusters by single-dish and interferometric experiments (for
a review see \cite{carlstrom_2002}) it has not not yet been
convincingly detected in any galaxy groups. However, a new generation
of high resolution CMB experiments with superior detector sensitivity
offers the hope of detecting and constraining the properties of the hot
gas in galaxy groups.

The outline of this paper is as follows. In Section 2 we construct
models of the hot gas in virialized halos of a given mass and
redshift, over the mass range $M_{\rm vir}=10^{13} M_\odot - 10^{15} M_\odot$
and redshift range $z=0 - 1.5$.  The models take into account
constraints on the temperature, entropy injection and gas fraction
from X-ray observations. In Section 3 we utilise multi-frequency
filtering techniques to determine the detectability of hot gas in
nearby halos with current and forthcoming microwave background
experiments such as the Atacama Cosmology Telescope (ACT~
\cite{act_kosowsky_2006}), the South Pole Telescope (SPT~ \cite{spt_ruhl_2004})
and the PLANCK mission \cite{planck_collab_2006}. In a final section we
summarize our findings and discuss how detection of the SZ effect in
group halos will constrain gas models and allow a
measurement of physical parameters such as the entropy injection and
baryon fraction in these halos. Throughout this paper we assume a flat
$\Lambda$CDM cosmological model with parameters $h=0.72,~\Omega_m=0.26, 
~\Omega_b=0.044, ~\Omega_{\rm de}=0.74, ~w_{\rm de}=-1.0,$ and $\sigma_8 = 0.8$ 
that provide a best fit to the WMAP 5 year data \cite{dunkley_2008}.

\section{Hot gas halo models}

Historically the isothermal $\beta$ model \cite{Cavaliere_1976} has
been used to describe the spatial distribution of hot gas in galaxy
clusters, primarily to model the X-ray emission originating from
thermal bremsstrahlung of the hot intracluster gas
\cite{Sarazin_1986}. The $\beta$ model provides a convenient
analytical form that has been popular for X-ray surface brightness
profile fitting.  However, over the past decade it has been realised
that X-ray observations of density profiles at large radii do not
favour the $\beta$ model: the isothermal model provides a poor
description of the temperature profiles of intracluster gas, with
observed temperature profiles declining at large radii and cooling
flows observed in the central regions of some clusters
\cite{Vikhlinin_2006}. Furthermore, it has been shown that there is a fundamental
incompatibility between the $\beta$ model parameters fitted using X-ray data, and
those fitted using data based on the SZ effect \cite{Hallman_2007},
indicating that the simple $\beta$ model is not sufficiently realistic
to describe the observed cluster gas physics.

In this paper, we will study two different models for the hot gas in
dark matter halos.  Our first model, which we will refer to as the
polytropic model, assumes that the hot gas follows a polytropic
equation of state and traces the dark matter in the outskirts of the
halo (e.g.~\cite{ks_2001, ks_2002}). As discussed in \citet{ks_2001}
this model is in good agreement with observed X-ray surface brightness
profiles and the mass-temperature relation above temperatures of a few
$\mbox{keV}$.  Our second model, which we refer to as the entropy
model, builds upon the first and attempts to account for
non-gravitational feedback by adding entropy to the hot gas, similar
to \citet{Voit_2002}.  In both cases, the hot gas is in hydrostatic
equilibrium with the underlying dark matter potential.

\subsection{Dark matter halo}

We define a dark matter halo at redshift $z$ with virial mass $\Mvir$
and radius $\Rvir$ to have a characteristic average density equal to
$\Delta_{\rm vir}(z)$ times the critical density,
\begin{equation}
  \rho_c(z) = \frac{3H^2(z)}{8\pi G}\ .
\end{equation} 
The Hubble parameter relative to its present value is given by
\begin{equation} {H(z) \over H_0} = \left[\Omega_m (1+z)^3 +
    \Omega_{\rm de}(1+z)^{3(1+w_{\rm de})}\right]^{1/2}\ ,
\end{equation} 
where $\Omega_{\rm m}$ and $\Omega_{\rm de}$ represent the density of
the matter and dark energy components, respectively, relative to the
critical density today.

In the spherical collapse model \cite{peebles_1980} for a SCDM
cosmology with $\Omega_{m} = 1$ and $\Omega_{de} = 0$, the virial
collapse factor has a constant value $\Delta_{\rm
  vir}(z)=18\pi^2$. Using numerical simulations, \citet{bn_1998} found
for $\Lambda$CDM cosmologies that the parametric form
\begin{equation}
  \Delta_{\rm vir}(z)= 18\pi^2 + 82 [\Omega(z) -1] -39 [\Omega(z) -1]^2\ ,
\end{equation} 
where
\begin{equation}
  \Omega(z) \equiv \Omega_{m} (1+z)^3 \left[{H_0 \over H(z)}\right]^2\ ,
\end{equation}
provides a good fit over a wide range of $\Lambda$CDM cosmological
models. Once the characteristic average density is chosen, the virial
mass and radius become uniquely related.

We assume that the dark matter density follows the self-similar NFW
profile \cite{nfw_1997},
\begin{equation} \rho_{dm}(x) = \frac{\rho_s}{x(1+x)^{2}},\qquad
  x\equiv r/r_s\ ,
\end{equation} 
where $r_s$ is the scale radius of the halo. By integrating the
density profile and equating the mass found within the virial radius
to $M_{\rm vir}$, we obtain the density normalization,
\begin{equation}
  \rho_s=\frac{\Mvir~c^3}{4\pi\Rvir^3 ~m(c)}\ , 
\end{equation}
where the function,
\begin{equation}
  m(x) = \ln(1+x)-\frac{x}{1+x}\ ,
\end{equation}
is the dimensionless dark matter mass profile. The concentration
parameter is defined as $c=\Rvir/{r_s}$ and we adopt the
fitting formula,
\begin{equation}
  c(\Mvir,z) = \frac{7.5}{1+z} \left[\frac{\Mvir}{\Mnl}\right]^{-0.1}\ ,
\end{equation}
to describe the dependence on virial mass and redshift following
\citet{dolag_2004}. The dependence on cosmological parameters is
captured by scaling the virial mass by the nonlinear mass
$\Mnl$. In the spherical collapse model, the nonlinear mass is
defined as the mass $M=(4/3)\pi\bar{\rho}R^3$ enclosed within a sphere
of radius $R$ for which the variance of the linear density field
$\delta$, smoothed by a tophat filter, equals the square of the
critical overdensity threshold $\delta_c=1.68$.  For the chosen
cosmology, the nonlinear mass is $\Mnl=2.818\times 10^{12}
h^{-1}M_\odot$ at redshift $z=0$.

The exact dependence of the concentration parameter on mass and
redshift is still uncertain as various parametrizations have been used
in the literature (e.g. \cite{nfw_1997, seljak_2000, bullock_2001}).
However, as previously observed \cite{ks_2002}, we find that our
results are insensitive to the exact choice of the concentration
parameter. This is due to the fact that the SZ effect is less
sensitive to the central regions, where the changes in the
concentration parameter are most important.

During the formation of virialized halos, the collisionless dark
matter undergoes violent relaxation and the collisional baryons get
shockheated.  According to the virial theorem, the internal energy of
a virialized halo is twice its gravitational potential energy.  From
this relation, we can define a characteristic velocity dispersion for
the dark matter and a virial temperature, \bea
T_{\rm vir} & = & {1\over 3} G\,\mu\,m_p\,\frac{\Mvir}{\Rvir} \\
& = & 0.034 \left({\Mvir \over 10^{12}M_\odot}\right)^{2/3} \left({
    \Delta_{\rm vir}(z) \over 100}\right)^{1/3} \left({H(z) \over
    H_0}\right)^{2/3} \mbox{keV}\ , \nonumber \eea for the shockheated
gas.  We assume a fully ionized gas with a mean molecular weight $\mu
= {4\over 3+5X_H}=0.588$ for a hydrogen mass fraction of $X_H=0.76$.

\subsection{Polytropic model}

We first consider a model where the gas follows a polytropic equation
of state, $P_g\propto\rho_g^\gamma$, with polytropic index $\gamma$.
The gas density and temperature profiles can be parametrized as, \bea
\rho_g(x) & = & \rho_{g,0} \ypoly(x)\ ,\\[10pt]
T_g(x) & = & T_{g,0} \ypoly^{\gamma-1}(x)\ , \eea where $\ypoly(x)$ is
the dimensionless gas density profile and the coefficients
$\rho_{g,0}\equiv\rho_g(0)$ and $T_{g,0}\equiv T_g(0)$ are two
boundary conditions.  Assuming that the gas is in hydrostatic
equilibrium with an NFW potential, we obtain the analytical solution
(e.g. \cite{ks_2001}), \bea
\ypoly(x) & = & \left[1-B f(x)\right]^{1/(\gamma-1)}\ , \nonumber\\[10pt]
f(x) & = & 1 - \frac{\ln(1+x)}{x}\ , \\[10pt]
B & = & \left(\frac{\gamma-1}{\gamma}\right)
\left(\frac{3\Tvir}{T_0}\right)\left[ \frac{c}{m(c)} \right]\ . \nonumber \eea

For a given mass and redshift, there are 3 free parameters: the
polytropic index $\gamma$, the central gas temperature $T_{g,0}$, and
the central gas density $\rho_{g,0}$.  In \citet{ks_2001,ks_2002} these
parameters were specified by requiring that the gas density profile
matches the dark matter density profile in the outer parts of the
halo.  This assumption is known to be in good agreement with adiabatic
hydrodynamic simulations, but it remains unclear how valid it is for
radiative simulations which include cooling, star formation, and
feedback.  We will take an alternative approach, choosing appropriate
values for the free parameters such that they are in agreement with
hydrodynamic simulations, are complimentary with other recent
semi-analytical models, and are flexible enough for us to apply them
to our non-polytropic model.

We set the polytropic index to be $\gamma=1.2$ as suggested by
hydrodynamic simulations and used in other recent semi-analytical
models (see \cite{ostriker_2005}, and references therein).  This value
is also consistent with the range of values used by
\citet{ks_2001,ks_2002}, who parametrized the polytropic index as a
function of the concentration parameter and found that it varies only
weakly with $c$.  In reality, the effective polytropic index
$\geff\equiv {\rm d}\ln P_g/{\rm d}\ln\rho_g$ is likely to be a
function of radius.  Towards the center of the halo where cooling is
more efficient because of the higher densities, the temperature can
decrease, resulting in $\geff<1$.  In the outskirts of the halo beyond
the virial radius where shockheating is less efficient, the effective
polytropic index will approach the characteristic value $\geff=1.62$
for the IGM \cite{hg_1997}.  Furthermore, non-gravitational feedback
can also change the equation of state. In our second model, we
consider a profile where the effective polytropic index is scale
dependent.

Halos found in adiabatic hydrodynamic simulations are generally well
described by the virial theorem.  The temperature at the virial radius
is close to the virial temperature and the central and average
temperature are found to be slightly higher (e.g. \cite{frenk_1999,
  rasia_2004}).  Therefore, we choose to equate the temperature at the
virial radius, $T_{g,c}\equiv T_g(c)$, to $\Tvir$ and this fixes the
central temperature as,
\begin{equation}
  T_{g,0}=T_{\rm vir}+3\Tvir\left(\frac{\gamma-1}{\gamma}\right)\left[\frac{cf(c)}{m(c)}\right]\ .
\end{equation}
Our chosen values for $T_{g,0}$ and $T_{g,c}$ are also consistent with
the range of values used by \cite{ks_2001,ks_2002}.

We normalize the density profile by fixing the mass of hot gas within
the virial radius, $\Rvir,$ and compare the gas fractions predicted by this model
to X-ray constraints within $r_{500}$.  X-ray observations of hot clusters
have demonstrated that the cumulative gas fraction within $r_{500}$
approaches a constant value that lies in the range $f_{\rm gas,500}
\approx 0.10$ to $0.16$ independent of cluster mass (\cite{Sun_2008,LaRoque_2006}). 
Furthermore there is good evidence that $f_{\rm
  gas}(<r_{200})$ converges to the universal baryon fraction $f_b$
(\cite{Vikhlinin_2006,Sanderson_2003,McCarthy_2006}). These results agree with cosmological
hydrodynamic simulations, which also suggest that for massive
clusters, there is very little evolution of the gas fraction with
redshift within $\Rvir$ \cite{Kay_2004,Eke_1998,Ettori_2004,Kravtsov_2005,Kay_2006}.
  
Observational constraints on the gas fraction in the cooler halos of
galaxies and groups of galaxies are much weaker than the cluster
measurements. However, there is evidence for a decrease of $f_{\rm
  gas}(<r_{500})$ with decreasing halo mass. It also appears that
$f_{\rm gas}(<r_{200})$ does not approach the universal baryon
fraction $f_b$ (\cite{Sanderson_2003}, see also
\cite{McCarthy_2006}). The main process that changes the gas fraction
while maintaining the baryon fraction is condensation of cold, low
entropy gas into stars. The lower value of $f_{\rm gas}$ for smaller
halos is consistent with the higher stellar fraction measured for
lower mass halos (\cite{Lin_2003,Gonzalez_2007}) combined with
the fact that lower mass halos are more sensitive to non-gravitational
heating which can expel gas from their shallower potentials. There are
few observational constraints on the evolution of the gas fraction
with redshift for cooler systems, but simulations suggest that there
is little evolution of the baryon fraction from $z=1$ to $z=0$ down to
the galactic scales of $M_{\rm vir}\simeq 10^{12}M_\odot$
\cite{Crain_2006}.

\begin{figure}[!t]
  \includegraphics[width=0.5\textwidth]{\plotsdir 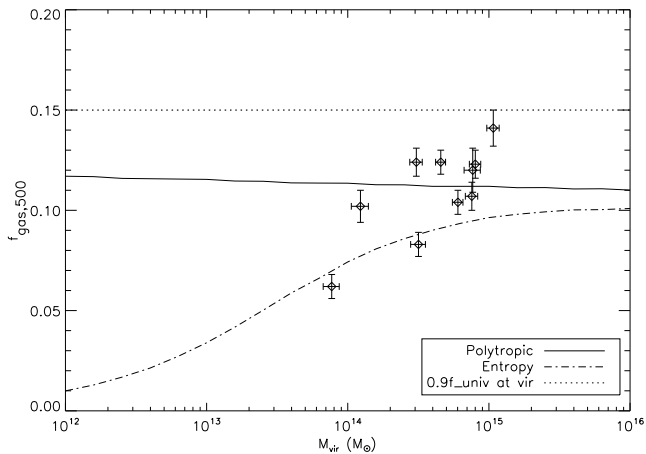}
  \caption{Integrated gas fraction within $r_{500}$ for the polytropic (solid curve) and entropy injection (dot-dashed curve) models, each normalized as described in the text. The model curves are compared to the universal baryon fraction, as measured from the WMAP 5 year data, with 10\% removed to account for stars (dotted curve). The
data points were obtained by \citet{Vikhlinin_2006} from a CHANDRA sample of nearby relaxed 
galaxy clusters.} \label{fig:f500}
\end{figure}

Taking these uncertainties into account, we assume that the baryon
fraction within the virial radius is given by the cosmic fraction
$\Omega_b/\Omega_m$, but allow for a fraction $f_*= 0.1$ of
baryons in the form of stars.  We also assume that $f_{\rm gas, vir}$
is redshift-independent (for \mbox{$z\lta 1$}) for all halo masses
($M_{\rm vir} = 10^{13} - 10^{15} M_\odot $) that we consider. The
central gas density is then given by,
\begin{equation}
  \rho_{g,0} = \frac{(1-f_*)\frac{\Omega_b}{\Omega_m}\Mvir}{4\pi r_s^3\int_0^c \ypoly(x)x^2 dx}\,.
\end{equation}
Note that \citet{ks_2001, ks_2002} chose to fix the gas fraction at the virial
radius to the cosmic value, without any explicit allowance for stars.
In general, the integrated gas mass and pressure within the virial
radius from our model agrees well with theirs with typical differences
of order ten percent.  We note that this simple
parametrization does indeed fit X-ray observations as can be seen from
Fig.~\ref{fig:f500}.

Now that the density and temperature of the gas are completely
specified for the polytropic model, we can write down additional
quantities that are relevant to SZ and X-ray observations.  The
electron pressure and entropy are given as,
\begin{equation}
P_e = n_e k T_e = P_{e,0}\ypoly\ ,
\end{equation}
and
\begin{equation}
S_e = T_e n_e^{-2/3} = S_{e,0}\ypoly^{\gamma-5/3}\ ,
\end{equation}
where $P_{e,0}\equiv n_{e,0}kT_{e,0}$ and $S_{e,0}\equiv
T_{e,0}n_{e,0}^{-2/3}$ are the central electron pressure and entropy,
respectively.  The electron number density $n_e$ is calculated from
the gas density assuming a fully ionized gas and the electron
temperature is assumed to be equal to the gas temperature.

\subsection{Entropy model}

Observations indicate that in the inner regions of low-temperature
clusters there is excess entropy above the predictions of the
self-similar model (\cite{Ponman_1999,Lloyd-Davies_2000,Finoguenov_2006}).  
Together with the observed departure from
the simple scaling relations suggested by purely gravitational physics
\cite{Arnaud_1999}, this means that non-gravitational effects should
be included when modeling the ICM gas distribution.  Various authors
have suggested that the ICM was heated by some energy input, e.g. via
star formation, SN explosions or AGN feedback. For example, \citet{Voit_2003}
showed that preheating can explain the entropy profiles of
groups. This model was also studied \cite{reid_2006} in the context
of SZ observations. \citet{ostriker_2005} constructed a model including
various non-gravitational processes that could be used to match the
observed scaling relations of clusters (see also
\cite{Bode_2006}).

Similar to \citet{Voit_2002} (see also \cite{Balogh_2005} and
\cite{Younger_2007}), we allow for an additive term ${S_{\rm
    inj}}$ to the polytropic entropy profile that incorporates the
combined effect of non-gravitational processes, such as feedback from
supernovae and galactic nuclei, and radiative cooling and star
formation.  ${S_{\rm inj}}$ is constant with radius so that
\begin{equation}
  S=S_0\,y^{\gamma-5/3}_{\rm ent}+S_{\rm inj}\,,
\end{equation}
The addition of an entropy term modifies the temperature profile
according to
\begin{equation}
  T_g(x)=T_{g,1}~y_{\rm ent}^{\gamma-1}(x)+T_{\rm inj}~y_{\rm
    ent}^{2/3}\, ,
\end{equation}
so that the central temperature is now $T_g(0)=T_{g,1} +
T_{\rm inj}$.  Here $T_{\rm inj}$ is the amount of injected thermal
energy per particle, and relates to the amount of injected entropy via
\begin{equation} {S_{\rm inj}}=100 \left(\frac{T_{\rm inj}}{1 {\rm
        keV}}\right) \left(\frac{n_{e0}}{10^{-3}{\rm
        cm}^{-3}}\right)^{-2/3} ~\mbox{keV cm}^2.
\end{equation} 

The form of heating proposed here is effective at increasing the
temperature in the inner regions of the halo but has little effect in
the outskirts of the halo for all but the lowest mass halos (see
Fig.~\ref{fig:temperature_profiles}).  The additional entropy due to
the non-gravitational heating term breaks the self-similarity of the
cluster physics. This very simple model provides a good
phenomenological description that is adequate for our purposes in
spite of its shortcomings.
  
We assume that the gas remains in hydrostatic equilibrium with the
dark matter, which yields an implicit equation for the
modified profile $y_{\rm ent}$, which we solve numerically.
For a given mass and redshift, there are 4 free parameters in this
model: the index $\gamma$ that characterizes the polytropic part of
the temperature, the central gas temperature, the central gas density
and the amount of injected energy $T_{\rm inj}$.  In general the gas
will not retain a polytropic equation of state and its profile will be
altered. However, we want to continue to associate the polytropic-like
terms with virialization and shockheating, thus we keep $\gamma=1.2$
and normalize the polytropic term in the temperature equation just
like in the polytropic model. That means we set the value at the
virial radius to be equal to the virial temperature, which fixes the
constant $T_{g,1}$ via
\begin{equation}
  \label{t1,ent}
  T_{g,1}=\frac{T_{\rm vir}}{y_{\rm ent}^{\gamma-1}(c)}.
\end{equation}
Note that we implicitly assume
that the gas has first settled in the gravitational potential set up
by the dark matter before being redistributed by the injection of
entropy.

We normalise the gas density by assuming that the gas pressure at the
virial radius remains unchanged by the entropy injection, i.e. $P_{\rm
gas}^{\rm ent}(\Rvir)=P_{\rm gas}^{\rm poly}(r_{\rm
vir})$. This means that we assume that the gas settles back into a
pressure balanced hydrostatic equilibrium at the virial radius after
the energy injection. Note that with our choices, the entropy model
reduces to the polytropic model when $T_{\rm inj}$ goes to zero.
The resulting electron temperature, density, pressure and entropy profiles
for the polytropic model and entropy injection model are shown in
Figs.~\ref{fig:temperature_profiles}-\ref{fig:entropy_profiles} in the
Appendix, where they are discussed in more detail.

\begin{figure}[!t]
  \includegraphics[width=0.5\textwidth]{\plotsdir
    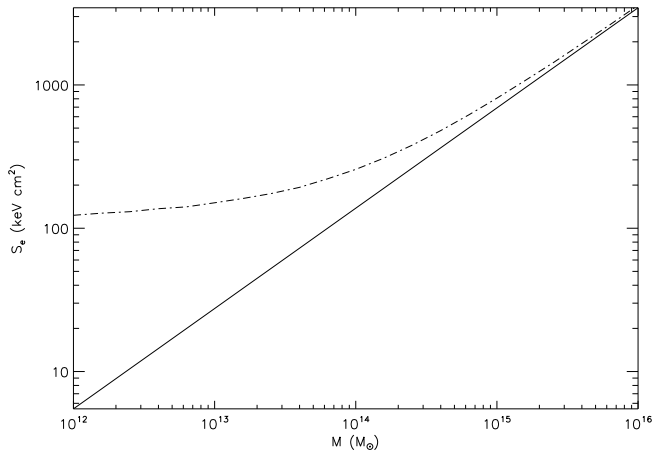}
  \caption{Entropy at $0.1\,\Rvir$ against halo mass, $M_{\rm vir}/M_\odot,$ for
    the polytropic model (solid curve) and the entropy model (dot-dashed curve), 
    showing a break in the scaling of the core entropy scaling as suggested by X-ray observations.}
  \label{fig:entropy_floor}
\end{figure}

In order to adjust the amount of heating $T_{\rm inj}$, we construct a
fitting function for $T_{\rm inj}$ such that
$S_{\rm inj} \approx 100~\mbox{keV}\,\mbox{cm}^2$ independent of halo mass
and redshift. It has been shown that this form and amount of feedback
reproduces the observed scaling relations of clusters
\cite{McCarthy_2004} (see \cite{Voit_2005} for a detailed discussion
of the feedback energy available from physical processes such as
supernovae or AGN heating).

In Fig.~\ref{fig:entropy_floor} we plot the entropy, $S_e$ at
$0.1\,\Rvir$ against virial mass for the different models. It can
be seen that the entropy model has more central entropy in
group halos.  This results from the feedback that heats the gas and
flattens the density profile, by pushing more gas from the center to
the outskirts, thereby breaking the scale invariance of the
entropy-mass relation. We also note that the level of feedback chosen in our entropy
injection model provides an entropy floor of $S\approx
100-200\,\mbox{keV}\,\mbox{cm}^2$ which agrees with entropy profiles
derived from X-ray observations
(see \cite{Ponman_1999,Lloyd-Davies_2000}) and simulations 
(see \cite{Finoguenov_2003}) of galaxy clusters and groups.

An additional constraint on the entropy injection parameter comes from
the X-ray luminosity-temperature relation. The X-ray luminosity within
radius $r_X$ is
\begin{eqnarray}
  L_X (<r_X) = && 
  1.41 \times 10^{35} \,\mbox{erg s$^{-1}$} \left(n_{e0} \over 10^{-3}\mbox{cm}^{-3}\right)^{2}
  \left( T_{e0} \over \mbox{keV}\right)^{1/2}   \nonumber\\ && \times
  \left( r_s \over \mbox{kpc}\right)^{3} \times 
  4 \pi \int^{r_X/r_s}_{0} {\cal Y}_X(x_p) ~x_p~ {dx_p}
  \nonumber\\
\end{eqnarray}
where the projected radius is $r_p=x_p\,r_s= \sqrt{r^2 - l^2\,r_s^2}$.
In the case of the entropy model the integrand ${\cal Y}_X$ is given
in terms of $y_{\rm gas}$, the dimensionless gas profile, as
\begin{eqnarray} {\cal Y}_X(r)= \sqrt{y_{\rm gas}^{\gamma+3}(r) +
    (T_{\rm inj}/T_{e0}) y_{\rm gas}^{14/ 3}(r)}\nonumber
\end{eqnarray}
while in the case of the polytropic model ${\cal Y}_X$ is simply
obtained by setting $T_{\rm inj} = 0$ in the above expression.

\begin{figure}[!t]
  \includegraphics[width=0.5\textwidth]{\plotsdir
  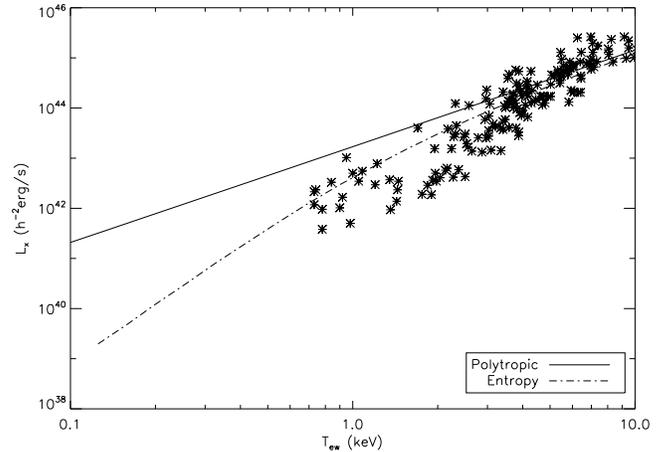}
  \caption{Integrated X-ray luminosity within $r_{500}$ against emission-weighted gas
    temperature, $T_{\rm ew},$  in the polytropic model (solid curve) 
    and the entropy model (dot-dashed curve). The data points were obtained by \citet{McCarthy_2004}
	from a compilation of CHANDRA and XMM-NEWTON data.}
  \label{fig:L_T_relation}
\end{figure}

In Fig.~\ref{fig:L_T_relation} we plot the integrated X-ray luminosity
within $r_{500}$ against the emission weighted electron temperature.
We see that the entropy injection model has significantly lower X-ray
luminosity for group halos, resulting in the observed
break of the scale invariant $L_X-T_X$ relation on group scales.

\section{Multi-frequency filtering of the halo SZ signal}

We now study the significance with which the halo SZ signal from
nearby groups and galaxies can be detected with forthcoming
multi-frequency CMB experiments.  Microwave maps contain not only the
thermal SZ, but a host of contaminants including primary CMB
anisotropies, kinetic SZ, microwave emission from galactic dust,
infrared and radio point sources, over and above the detector noise of
the experiment.  The importance of utilizing the SZ as a cosmological
probe has prompted several authors to develop specialized techniques
for detecting galaxy clusters through the SZ effect.  Proposed
techniques include the maximum entropy method \cite{hobson_1998},
fast independent component analysis \cite{maino_2002}, matched filter 
analysis \cite{herranz_2002b}, wavelet filtering \cite{pierpaoli_2005} 
and Wiener filtering \cite{tegmark_1996}. Map filtering, in
general, utilizes both the spatial and frequency information to
separate galactic foreground emission and extragalactic point source
contamination from the primary CMB and thermal SZ signals.

For the purposes of this investigation, which involves assessing the
level of detection of a SZ halo, we utilize a simple multi-frequency
Wiener filtering technique as described in \citet{tegmark_1996}
to separate the halo SZ signal from other components.  The method
allows us to determine the level of residual noise in the filtered
maps and a signal to noise ratio for each halo. Foreground and noise
subtraction is done in harmonic space which allows one to exploit the
fact that contaminants such as the CMB, galactic dust and extragalactic point sources
have power spectra that differ from that of the thermal SZ effect. 
We also include a contribution from the thermal SZ background that has the same 
frequency dependence as the halo SZ signal. The contamination arising from
the superposition of SZ sources along the line of sight has the potential to reduce
detectability and distort observable properties of the halo SZ signal \cite{holder_2007}.

\subsection{Multi-frequency Wiener filter}

We assume that we have microwave sky maps 
$d_i(\bfr)$ at pixel position $\bfr$ for ${\cal M}$ different frequencies. 
The signal in each map originates from ${\cal N}$ components $s_j(\bfr)$ such as the
primary CMB anisotropy, SZ sources, galactic foregrounds and extragalactic point sources,
so that $$\Delta T(\bfr, \nu) = \sum_j f_j(\nu) s_j(\bfr)$$ where 
$f_j(\nu)$ is the frequency dependence
of the $j$th component. In addition to these components 
each map contains detector noise $n_i(\bfr)$ which
we treat as random in each pixel. 
Then we can write our observation
as $$d_i(\bfr) = F_{ij} s_j(\bfr) + n_i(\bfr) $$ 
where the ${ \cal M} \times {\cal N}$ frequency
response matrix $F$ is defined as $F_{ij} = \int w_i(\nu) f_j(\nu) d\nu$. 
It is convenient to absorb
the beam response factor into the definition of the pixel noise \cite{knox_1995}
which allows us to set the response coefficients $w_i$ to unity.

We assume that the noise has zero mean $\langle n_i\rangle = 0$ with covariance 
$$\langle \tilde{n}_i(\bfl) \tilde{n}_j^{*}(\bfl') \rangle 
= (2\pi)^2 \delta(\bfl - \bfl')\tilde{N}_{ij}(l).$$  
We will consider the case of 
white noise, for which $\tilde{N}_{ij}({\bfl})$ is constant.
We assume that the signal and foreground components have means 
$$\langle s_i(\bfr)\rangle = A_i$$ with covariance 
$$\langle (\tilde{s}_i(\bfl) - {A}_i) (\tilde{s}_j^{*}(\bfl') - {A}_j)\rangle 
= (2\pi)^2 \delta(\bfl - \bfl')\tilde{S}_{ij}(\ell).$$
The condition that the signal and noise components are uncorrelated ensures that
the signal covariance matrix, 
$ {\tilde{S}}_{ij}(\ell) = \delta_{ij} \tilde{C}_{\ell,(j)},$
and noise covariance matrix,
$ {\tilde{N}}_{ij}(\ell) = \delta_{ij} \tilde{N}_{\ell,(j)},$ are diagonal. In 
what follows we will drop the tilde on harmonic space quantities.

The most general linear estimator $\hat{\bfs}$ of the signal can be
constructed from the data $\bfd$ as $$\hat{\bfs}(\bfr) = \int
W(\bfr-\bfr') \bfd(\bfr') d^2\bfr',$$ 
where $W,$ is an ${\cal N}\times {\cal M}$ weight matrix. We use
the flat sky approximation and work in harmonic coordinates. 
Some signals e.g., the primordial CMB, have a zero mean,
$A_i=0$. For a signal with non-zero mean, the condition of an unbiased estimator
requires that 
$$ \langle \hat{s}_i(\bfr) \rangle = 
\int ~ \sum_{j=1}^{\cal M} 
{W}_{ij}(\bfl) ~ {s}_j(\bfl)~ d^2\bfl=A_i,$$ where we have used Parseval's theorem.

The residual error in the maps from the noise and foregrounds is given by 
\begin{eqnarray} 
\langle {\Delta}_i(\bfr)^2\rangle = \langle (\hat{s}_i(\bfr) - {s}_i(\bfr))^2\rangle =  
\int \Delta_{i}^2(\ell) ~ d^2\bfl.
\end{eqnarray}
where $$\Delta_{i}^2(\ell) =  \sum_{j=1}^{\cal N} |(W_\ell F - I)_{ij}|^2 S_{\ell,j} + 
\sum_{j=1}^{\cal M} |W_{\ell,ij}|^2 N_{\ell,j} .$$
The first term accounts for the contamination of the desired signal by other components 
while the second term measures the detector noise. A non-zero mixing arises when two 
or more components have similar frequency dependence.
Requiring that the residual error is minimised we derive the Wiener filter weights 
$${W}_\ell = {S}_\ell F^T [F {S}_\ell F^T + {N}_\ell]^{-1}.$$

To set the threshold for detection we 
compare the mean SZ signal to the residual noise and define the signal-to-noise ratio as
$$ {\rm S / N}= { \langle \hat{s}_{\rm sz}(\bfr) \rangle \over 
\langle {\Delta}_{\rm sz}(\bfr)^2\rangle^{1/2} }= 
{ \int ~ \hat{s}_{sz}(\ell)~ \ell\, d\ell \over
\left[{\int \Delta_{\rm sz}^2(\ell) ~ \ell \, d\ell}\right]^{1/2}},$$
where the
recovered signal is given by 
$$\hat{s}_{sz}(\ell)= \sum_{j=1}^{\cal M} W_{{\rm sz},j}(\ell) ~ {s}_{sz}(\ell).$$
The signal to noise ratio depends on the range of $\ell$ over which the integration
is performed. We chose $\ell_{min}$ and $\ell_{max}$ to give the maximum signal to noise 
over this $\ell$ range -- in practice this would correspond to applying the appropriate 
high pass and low pass filters to the recovered map.

\begin{figure}[!t]
\includegraphics[width=0.45\textwidth]{\plotsdir 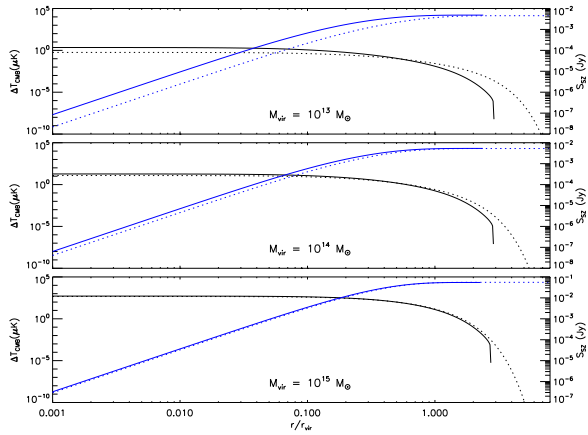}
\caption{SZ temperature distortion profile (black curves, decreasing from left to right) and integrated flux profile (blue curves, increasing from left to right) plotted against $r/\Rvir$ for the polytropic model (solid curves) and the entropy model (dotted curves). The panels from top to bottom show the SZ distortion 
and integrated flux profiles for halo masses $M_{\rm vir}=10^{13},10^{14}$ and $10^{15}~M_\odot$ respectively.
The sharp fall-off that is visible in the temperature distortion profile for the polytropic model 
is a result of the Gaussian smoothing described in the text, and occurs at larger radius for the
entropy injection model.}
\label{fig:sz_temperature_flux}
\end{figure}

\subsection{Halo Sunyaev-Zel'dovich Signal}

The upscattering of microwave background photons by hot electrons in the halo results in 
a projected Compton profile
\begin{equation}
y_{\rm comp}(r_p) = y_{\rm c0} \times 2 \int^{r_{\rm max}/r_s}_{0} 
\limits {\cal Y}_{\rm SZ}(r) dl
\end{equation}
where the central Compton distortion is
\begin{equation}   
y_{\rm c0}=8.0 \times 10 ^{-3} 
        \left(n_{e0} \over 10^{-3}\mbox{cm}^{-3}\right)
        \left( T_{e0} \over \mbox{keV}\right)
        \left( r_s \over \mbox{Mpc}\right) 
\end{equation}
and the projected radius is $r_p=x_p\,r_s= \sqrt{r^2 - l^2\,r_s^2}$.
In the case of the entropy model the projected pressure is given by
\begin{eqnarray}
{\cal Y}_{\rm SZ}(r)= \sqrt{y_{\rm gas}^{\gamma}(r) + (T_{\rm inj}/T_{e0}) y_{\rm gas}^{5/ 3}(r)}
\nonumber
\end{eqnarray}
where $y_{\rm gas}$ is the dimensionless gas profile, while in the
polytropic model the corresponding expression is obtained by setting
$T_{\rm inj}= 0$.

The resulting thermal SZ temperature distortion is given by 
\begin{equation}
  \Delta T(\theta) = j(x) ~T_{\rm cmb} ~y_{\rm comp}(\theta)
\end{equation}
where $\theta = r_p/d_A,$ the CMB temperature is $T_{\rm cmb}=2.73
~{\rm K}$ and the frequency dependence is given by
\begin{equation}   \label{eqn:sz_freq_dependence}
j(x) = {x(e^x+1) \over (e^x - 1)} - 4 ~, 
\qquad x = \nu /56.9~\mbox{GHz}.
\end{equation}
The projected temperature distortion profiles are shown in
Fig.~\ref{fig:sz_temperature_flux}. We note that the polytropic and entropy
injection models are almost indistinguishable for the most massive
halos. In Fig.~\ref{fig:sz_temperature_flux} we also show the cumulative
SZ flux within  radius $r_{\rm SZ}$, which is given by
\begin{equation}
S_{\rm SZ}(<r_{\rm SZ})= I_{\rm cmb} ~g(x) ~ \times 2 \pi \int^{r_{\rm SZ}/r_s}_{0} 
y_{\rm comp}(x_p) ~ x_p~  {dx_p}
\end{equation}
where
\begin{equation}   
g(x) = j(x)\, {x^4e^x \over (e^x-1)^2}, \qquad I_{\rm cmb} = 0.27 ~{\rm GJy}.
\end{equation}
While the SZ signatures of the largest clusters are nearly identical, it is 
clear from Fig.~\ref{fig:sz_temperature_flux} that 
the SZ temperature distortions of less massive group halos 
are sensitive to feedback effects even for models which have the same SZ flux at 
the virial radius.

\begin{figure}[t!]
  \includegraphics[width=0.5\textwidth]{\plotsdir 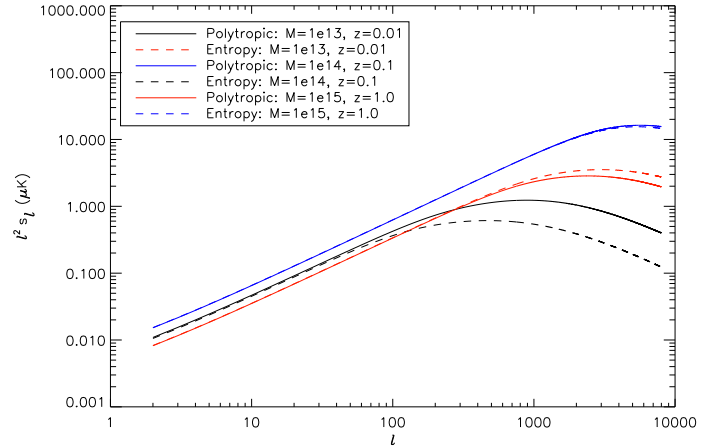}
  \caption{Halo spectra, $s_{\rm sz}(\ell),$ for the polytropic model (solid curves) 
		  and entropy injection model (dashed curves). The curves from top to bottom  
		show the spectra for halo masses $M_{\rm vir}=10^{13}, 10^{14}$ and $10^{15}~M_\odot$ 
		(blue, red and black) respectively.}
  \label{signal_vs_residual_3}
\end{figure}

The relevant quantities in harmonic space for determining the detectability
of a given SZ halo are the Bessel transform of the halo SZ distortion   
\begin{equation}   
  s_{\rm sz}(\ell) = 2\pi \, {\int_0}^{\theta_{\rm max}}\,  
  [\Delta T(\theta)/j(x)]\, J_0(\ell\,\theta)\, \theta \,d\theta 
\end{equation}   
where $J_0$ is the Bessel function of order zero, and the SZ halo power spectrum
\begin{equation}   
   C_{\ell,sz} = 2\pi \, {\int_0}^{\theta_{\rm max}}\,  
  [\Delta T(\theta)/j(x)]^2\, J_0(\ell\,\theta)\, \theta \,d\theta 
\end{equation}   
which enters the reconstruction error if the filter weights are not diagonal
in the SZ component i.e., $(WF)_{i\,sz}\neq \delta_{i\,sz}$. 

The sharp cutoff in the Bessel transform, combined with the fact that
the SZ distortion has not fallen to zero at $\theta_{\rm max},$ results in ringing
of the  profile spectrum and power spectrum in $\ell$-space. We therefore smooth
the density profile using a Gaussian profile so that 
$\rho_{gas}(r) \rightarrow \rho_{gas}(r) e^{-{r^2}\slash{\xi r_{\rm max}^2}}$ where $\xi$ is chosen
to ensure that the total gas mass is unchanged. 
The smoothed profile falls off sharply after $r_{\rm max}$ so in practice we integrate out
to $\theta_{\rm max}^{\rm smooth}$ that is a few times larger than $\theta_{\rm max},$ 
which suffices to remove the ringing. The resulting spectra have slightly
more ($\sim$ few percent) power on intermediate and large scales with the power
going smoothly to zero at large $\ell$.

In Fig.~\ref{signal_vs_residual_3} we compare the SZ halo spectrum $s_{\rm sz}(\ell)$ 
in the polytropic and entropy injection models for halos of mass
$M_{\rm vir}= 10^{13} M_\odot$ at redshift $z=0.01,$  mass $M_{\rm vir}= 10^{14} M_\odot$ 
at redshift $z=0.1,$ and mass $M_{\rm vir}= 10^{15} M_\odot$ at redshift $z=1.0$.
In the largest mass halo we note that both our gas models produce nearly identical 
spectra due to the fact that the heating is small relative to the thermal energy
of the hot gas. For the lower mass halos we observe that
the polytropic model produces a larger amplitude SZ signal due to
the higher density of gas in the central regions of the halo. The
temperature increase that occurs in the entropy injection models is
insufficient to compensate for the reduced density, which results in a
SZ profile with smaller amplitude for larger entropy injection. The trend is
monotonic suggesting that, for experiments with sufficient sensitivity
and frequency coverage to extract the SZ halo signal, the amount of
entropy injection may be measurable from these SZ observations.

\subsection{Foreground components}

We assume that the spatial and frequency dependence of each foreground
component can be written as a product
$$\Delta T_{(i)}^2(\nu,\ell) =  A_{(i)}^2 f_{(i)}^2(\nu) C_{\ell,(i)}~,$$
where $f_{(i)}(\nu)$ measures the average frequency dependence of a
component, $C_\ell$ is its spatial power spectrum and $A_{(i)}$

Note that $f_{(i)}(\nu)$ gives the frequency dependence of the
RMS fluctuations in thermodynamic temperature referenced to the CMB
blackbody.  We normalise the frequency term, $f_{(i)}(\nu),$ to be unity at
$\nu_*=56.9$~GHz and the spatial term $C_{\ell,(i)}$ to be unity at
$\ell_*=2$ so that the units are absorbed into the overall amplitude
$A_{(i)}$.  We now consider models for each of the foreground
components in turn.

\noindent
\subsubsection{Galactic dust emission}

We model the frequency dependence of thermal galactic dust emission as
\be 
f_{\rm dust}(\nu)=c(x)~c_{\ast}(x)~\frac{x_{\rm dust}^{3+\alpha}}{e^{x_{\rm dust}}-1},
\qquad x_{\rm dust} = h \nu / k_B T_{\rm dust}. 
\ee
where $\alpha$ is the emissivity index and
\be
c(x)=\left( \frac{2 \sinh \frac{x}{2}}{x}\right)^2, \quad
c_{\ast}(x)=\frac{1}{x^2}\frac{1}{2k}\left(\frac{hc}{k_B T_{cmb}} \right)^2
\ee
convert antenna temperature to thermodynamic temperature
and specific intensity to antenna temperature respectively.
In our model we assume an emissivity index $\alpha=$1.7 and a 
dust temperature $T_{dust}=$18~K \cite{tegmark_1999, draine_1999, ponthieu_2005}.

We model the spatial power spectrum of the thermal dust component
as a power law \be C_{l,{\rm dust}}=({\ell/\ell_*})^{-\beta}, \ee
where $\beta$ is the power law index. We set $\beta =
3$ which was the value derived from an analysis of the DIRBE maps
\cite{wright_1998}.  We fix the amplitude of the galactic dust
emission to be $A=10.2\,\mu$K at 56.9~GHz.

\begin{figure}[!t]
\includegraphics[width=0.5\textwidth]{\plotsdir 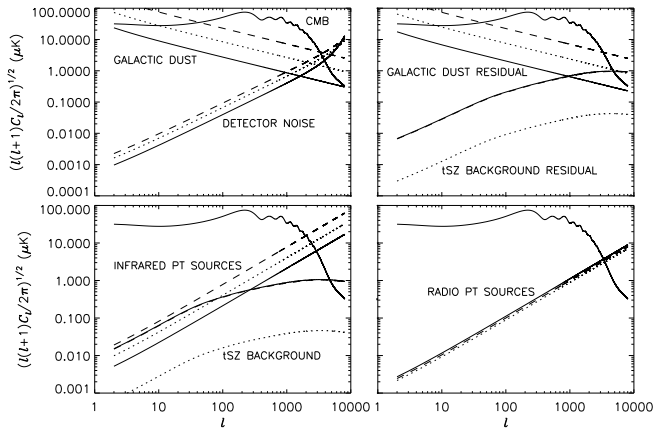}
\caption{Angular power spectra of the foreground components and ACT detector noise at 145 GHz (solid curves), 
         215 GHz (dotted curves) and 280 GHz (dashed curves) respectively. The top left panel displays 
   		the power spectra of the galactic dust, thermal SZ (tSZ) background and detector noise. 
		The top right panel displays the power spectra of the residual tSZ background and residual 
  		galactic dust. The bottom left and bottom right panels display the
  		spectra of infrared and radio point sources respectively. All spectra are 
  		compared to the lensed CMB spectrum (bold solid curve) in each panel.}
\label{fig:signals_plus_noise_all_channels}
\end{figure}

An analysis of the FIRAS and DIRBE datasets \cite{SFD_1998, FS_1999}
has provided evidence for two dust components with different 
temperatures and emissivities. We account for uncertainty in the emissivity 
by introducing a residual dust component with the same 
spatial power spectrum but with scatter, $\delta \alpha,$ in the emissivity index.  
We choose $\delta \alpha=0.3$ as
suggested by the analysis of \citet{FDS_1999}, which is consistent
with the results of \citet{draine_1999}.  The top left and top right panels
of Fig.~\ref{fig:signals_plus_noise_all_channels} display the power
spectra of the dust and residual dust components respectively.

\noindent
\subsubsection{Radio and infrared point sources}

We consider two point source populations: radio sources e.g., blazars,
and infrared point sources e.g., early dusty galaxies.  We model radio sources
using a fit to the WMAP Q-band ($\nu_o = 41$~GHz) data \cite{bennett_2003}: \be
\frac{dN}{dS_{\nu}}=\frac{N_o}{S_o} \left ( \frac{S_{\nu}}{S_o}
\right)^{-2.3} \ee where $N_o=$ 80$\, \mbox{deg}^{-2}$ and $S_o=$
1$\,$mJy.  Since we are mostly concerned with fluxes at the mJy level,
the slope of the distribution was altered to $-2.3$ from the fiducial
slope of $-2.7$ \cite{white_2004}. In the case of infrared point sources 
we use the fit \cite{borys_2003} to SCUBA observations \cite{holland_1999} 
at $\nu_o=350$~GHz, given by \be
\frac{dN}{dS_{\nu}} = \frac{N_o}{S_o} \left [
  \left(\frac{S_{\nu}}{S_o}\right) +
  \left(\frac{S_{\nu}}{S_o}\right)^{3.3} \right]^{-1}, \ee where
$N_o=$ 1.5 $\times 10^4 \, \mbox{deg}^{-2}$ and $S_o \simeq$
1.8$\,$mJy.
The sources fluxes were extrapolated to the frequencies of the various
CMB experiments using a power law, $S_\nu \propto \left(\nu/\nu_o
\right)^{\alpha}$.  We use a spectral index $\alpha = 0$ for radio
sources and $\alpha = 2.5$ for infrared sources \cite{white_2004}. 

The power spectra of the point sources is calculated from
\be C_{\textit{l}} = \left(\frac{dB}{dT}\right)^{-2} \int_0^{S_{cut}}
\frac{dN}{dS_\nu}S_\nu^2 dS_\nu, \ee where $dB/dT$ is the derivative of the Planck
spectrum and $S_{cut}$ is
the imposed flux cut, which we assume to be 5 mJy for ACT and SPT \cite{white_2004}
and 250 mJy for PLANCK \cite{vielva_2001}. We
assume that the point sources are spatially uncorrelated on the sky,
thus the power is constant at all multipoles. The power spectra of
infrared and radio point sources are displayed in the bottom left and bottom right panels
of Fig.~\ref{fig:signals_plus_noise_all_channels} respectively. 

\noindent
\subsubsection{Cosmic microwave background}

The cosmic microwave background anisotropy has a constant frequency
dependence with reference to the blackbody temperature so that $f_{\rm cmb}(\nu)= 1.$
The CMB power spectrum, $C_\ell$, was calculated using the CAMB software package 
\footnote{CAMB: http://www.camb.info} 
using the WMAP 5 year best fit cosmological model.
The lensed CMB angular power spectrum is shown in Fig.~\ref{fig:signals_plus_noise_all_channels}.  

\begin{table}
\begin{tabular}{|c|c|c|c|} \hline
Experiment & $\nu$ (GHz) & $\sigma_p$ ($\mu$K/pixel) & $\theta_b$ ($^\circ$)\\\hline
  &               100.0 & 4.5   & 0.18   \\
  &               143.0 & 5.5   & 0.13   \\
  PLANCK: all-sky &       217.0 & 11.8  & 0.092  \\
  &               353.0 & 39.3  & 0.083  \\
  &               545.0 & 401.3 & 0.083  \\ \hline
  &               145.0 & 2.0 (8.9)  & 0.028  \\
ACT: 200 deg$^2$ (4000 deg$^2$) & 215.0 & 5.2 (23.3)   & 0.018  \\
  &               280.0 & 8.8 (39.4)  & 0.015 \\  \hline
  &               95.0  & 9.1 (2.0)  & 0.026   \\
  &               150.0 & 13.4 (3.0)  & 0.017   \\
SPT: 4000 deg$^2$ (200 deg$^2$)& 219.0 & 41.2 (9.2) & 0.012  \\
  &               274.0 & 71.4 (16.0) & 0.009  \\
  &               345.0 & 583.9 (130.6) & 0.007  \\ \hline
\end{tabular}
\caption{Experimental specifications for the PLANCK \cite{planck_collab_2006}, ACT \cite{act_kosowsky_2006}
and SPT \cite{spt_ruhl_2004} experiments. In addition to the nominal ACT and SPT surveys, the specifications for 
a wider ACT survey and deeper SPT survey are also listed, where we have assumed a fixed total integration
time in rescaling the pixel noise.}
\label{table:exptspecs}
\end{table}

\noindent
\subsubsection{SZ background}

The projection of SZ sources of varying mass and redshift along the
line of sight creates a diffuse SZ background which can contaminate halo SZ
observables.  To account for the contamination of the
foreground halo signal by background clusters, we model the SZ
background statistically by using its power spectrum. Ideally one would
utilise a simulated map of SZ halos to study the contamination due
to projection effects but we defer this investigation to a future
publication. This map would also take into account the contamination from hot gas
outside collapsed structures, though \citet{hernandez-monteagudo_2006} have shown
that this component does not significantly contribute to the thermal SZ power spectrum.

The frequency dependence of the thermal SZ background is the same as that
of the thermal SZ halo signal given in \ref{eqn:sz_freq_dependence}. 
The power spectrum of the SZ background is computed following
\citet{ks_2002} over the mass range $10^{12} - 10^{16} M_\odot$.
We also allow for an uncertainty in the SZ background which we conservatively
model as arising from background halos smaller than $5\times 10^{14} M_\odot$.
The top left and top right panels of Fig.~\ref{fig:signals_plus_noise_all_channels} 
display the power spectra of the SZ background and residual SZ background.

\subsection{Detector noise and experimental specifications}

We model the detector noise as an additional sky signal
\cite{knox_1995} with power spectrum \be
N_{\ell,i}=w_{i}^{-1}e^{\theta_{b,i}^2 \ell(\ell+1)} \ee in a given
frequency band, $i$. In this case each of the sky signals is not convolved
with the experimental beam.  We assume that the experimental beam is
a Gaussian of width $\theta_{b,i}$ so that the full width at half
maximum is given by FWHM$_{i} \, =\sqrt{8\ln 2}~\theta_{b,i}$. The inverse
noise weight \be w_{i}^{-1} = \sigma_{p,i}^2 \times \theta_{b,i}^2 \ee is
defined as the noise variance per pixel times the pixel area in steradians.

We consider three nominal experiments, a shallow all sky SZ survey 
by the PLANCK surveyor
\footnote{PLANCK: http://www.rssd.esa.int/index.php?project=planck}, 
a 200 deg$^2$ SZ survey by the Atacama Cosmology Telescope (ACT) 
\footnote{ACT: http://www.physics.princeton.edu/act/} and a 
4000 deg$^2$ survey by the South Pole Telescope (SPT)
\footnote{SPT: http://pole.uchicago.edu/}. We also consider a
wider ACT survey (over 4000 deg$^2$) and a deeper SPT survey (over 200 deg$^2$)
where we have rescaled the pixel noise using a fixed total integration time.
Specifications for the various experiments are listed in Table~\ref{table:exptspecs} 
\cite{act_kosowsky_2006,planck_collab_2006,spt_ruhl_2004}.
The top left panel of Fig.~\ref{fig:signals_plus_noise_all_channels} displays the detector noise 
power spectra for the ACT experiment.

\begin{figure}[t!]
  \includegraphics[width=0.5\textwidth]{\plotsdir 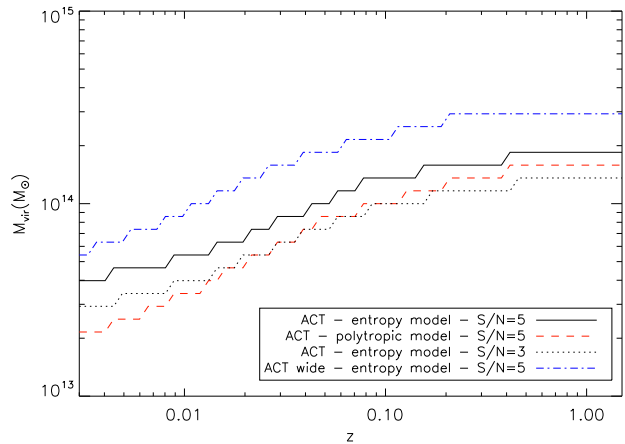}
  \caption{The minimum detectable mass, $M_{\rm vir},$ of SZ halos as a function of redshift, $z,$ for the ACT experiment. The minimum detectable mass is plotted for an entropy model with $S/N=5$ (solid curve) and $S/N=3$ (dotted curve), and a polytropic model with $S/N$=5 (dashed curve) for the nominal ACT survey. Also shown
is the minimum detectable mass with the ACT wide survey for an entropy model with $S/N=5$ (dot-dashed curve).}
  \label{halo_detection_limits_act}
\end{figure}

\section{Detectability of the halo SZ signal}

\begin{figure}[t!]
  \includegraphics[width=0.5\textwidth]{\plotsdir 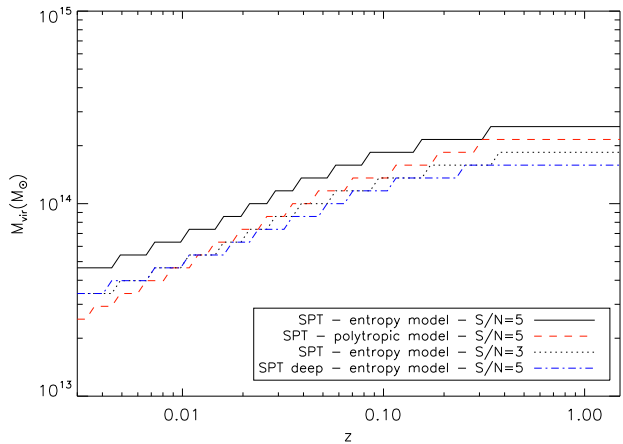}
  \caption{The minimum detectable mass, $M_{\rm vir},$ of SZ halos as a function of redshift, $z,$ for the SPT experiment. The minimum detectable mass is plotted for an entropy model with $S/N=5$ (solid curve) and $S/N=3$ (dotted curve), and a polytropic model with $S/N$=5 (dashed curve) for the nominal SPT survey. Also shown
is the minimum detectable mass with the SPT deep survey for an entropy model with $S/N=5$ (dot-dashed curve).}
  \label{halo_detection_limits_spt}
\end{figure}

\begin{figure}[t!]
  \includegraphics[width=0.5\textwidth]{\plotsdir 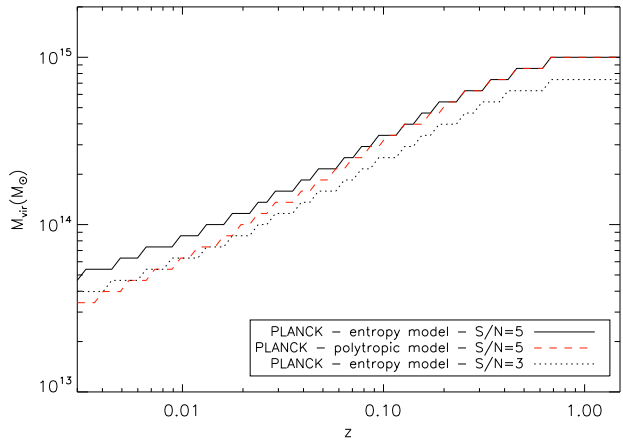}
  \caption{The minimum detectable mass, $M_{\rm vir},$ of SZ halos as a function of redshift, $z,$ for the PLANCK experiment. The minimum detectable mass is plotted for an entropy model with $S/N=5$ (solid curve) and $S/N=3$ (dotted curve), and a polytropic model with $S/N$=5 (dashed curve).}
  \label{halo_detection_limits_planck}
\end{figure}

We now study the detectability of SZ halos in our mass and
redshift range for the polytropic and entropy injection models presented above.
In Fig.~\ref{halo_detection_limits_act} we plot the minimum detectable halo mass, or threshold mass,
as a function of redshift for the ACT experiment using a detection significance of $S/N=3$ and $S/N=5$. Similar plots for the PLANCK and SPT experiments are presented in Fig.~\ref{halo_detection_limits_spt}
and Fig.~\ref{halo_detection_limits_planck} respectively.

At high redshift ($z\sim 1$) the ACT experiment reaches a threshold mass of $2\times10^{14} M_\odot$  
for a signal to noise ratio of five and sky coverage of $200~ \mbox{deg}^2,$ which is 
similar to the completeness limit presented in \cite{sehgal_2007}. The threshold mass at high redshift 
is similar for the SPT experiment but larger ($\sim 10^{15} M_\odot$) for the all-sky PLANCK survey which 
has larger pixel noise. At low redshift, however, it is interesting to note that the threshold mass 
for ACT, in the case of the entropy injection model, drops below $10^{14} M_\odot$ at $z\approx 0.05$ 
and as low as $4\times10^{13} M_\odot$ at $z\approx 0.005 \,(\approx 20 \,{\rm Mpc}),$ and is similar
in the case of the SPT-deep survey. Although the threshold mass at low redshift is higher for the SPT 
wide ($4000~ \mbox{deg}^2$) survey and the all-sky PLANCK survey it is still possible to detect
group-sized halos below $10^{14} M_\odot$ at $z<0.02$. This indicates that the thermal SZ effect in 
nearby group-sized halos can be detected with multi-frequency observations that reach pixel 
sensitivities of a few $\mu$K, and that the detectability of these halos can be improved with longer
integration times.

\begin{figure}[h!]
  \includegraphics[width=0.5\textwidth]{\plotsdir 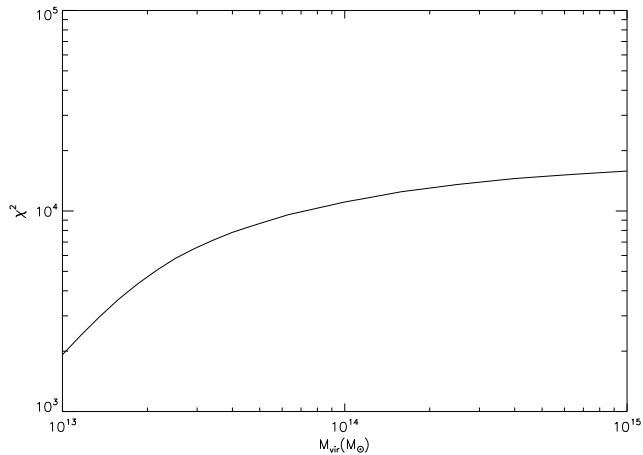}
  \caption{Significance with which the entropy model can be distinguished from the polytropic model by the nominal ACT survey as a function of halo mass, $M_{\rm vir}$. The $\chi^2$ statistic is calculated by 
comparing the difference between model spectra to the residual noise and foreground level of the ACT experiment.}
  \label{chisq_poly_vs_ent}
\end{figure}

For the polytropic model the detection levels are more optimistic because 
the SZ signal is larger due to the gas being more concentrated, however, as discussed
in the previous section this model is less realistic, particularly for low mass clusters
and groups. To determine the significance with which one can distinguish the entropy model
from the polytropic model via measurements of the SZ distortion in these halos we computed
the $\chi^2$ statistic that compares the difference in spectra between these models 
to the residual noise and foreground level from the nominal ACT experiment. In particular we were interested in the 
leverage one gains from the detection of galaxy groups and low mass clusters. In Fig.~\ref{chisq_poly_vs_ent} 
we plot the $\chi^2$ statistic as a function of virial mass. We observe that the information
gained from group halos is of the same order of magnitude as the information gained from
larger clusters because the larger differences in the SZ spectra, that results from the increased impact of the 
entropy injection in lower mass halos, compensates the larger residual noise. For group halos detected with higher significance these measurements will provide useful joint constraints on the gas fraction and level of entropy injection in these halos.

It is interesting to quantify the yield of galaxy groups and low mass clusters that are detectable in these surveys. We calculate the yield by integrating the cluster abundance from the minimum survey threshold mass to a cutoff mass of $2 \times 10^{14} M_\odot,$ which is roughly the minimum mass quoted for the detection of clusters in upcoming SZ cluster surveys e.g., \cite{sehgal_2007}. In choosing this cutoff mass our aim is to quantify the additional yield of SZ halos, over and above the yield of more massive clusters, in these surveys.
The anticipated number of detectable galaxy groups and low mass clusters are given in Table \ref{table:yield} for the different surveys. We note that the mass function is steep so the yield is very sensitive to the minimum and maximum mass limits, consequently the numbers quoted here should only be taken as a rough guide to the anticipated yields. We observe that all surveys will yield a reasonable number of detectable halos below the mass cutoff at the $S/N=5$ level, with the numbers increasing significantly for halos detected at the lower significance of $S/N =3$, though the contamination will be higher at this level. It is generally the case that the deep surveys produce a higher yield than the wide surveys for a given model, presumably because the mass function is so steep in this mass range. This trend is reversed in the case of the SPT wide survey at the $S/N=3$ level, however, because the 
increased sky area is sufficient to compensate for the reduced sensitivity. We also note that the yields for the polytropic model are larger than those for the entropy model due to the larger signal in the polytropic model. 

\begin{figure}[t!]
  \includegraphics[width=0.5\textwidth]{\plotsdir 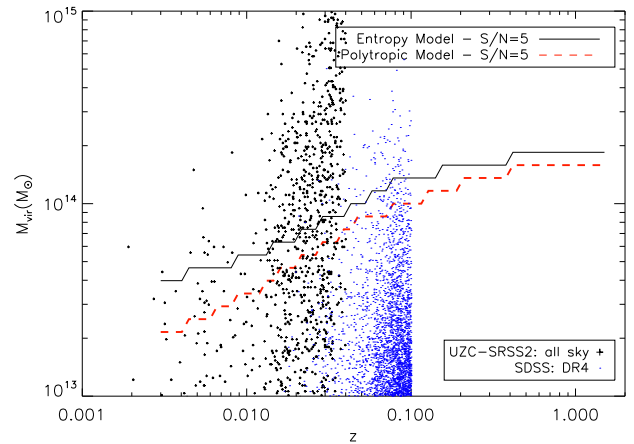}
  \caption{The minimum detectable mass, $M_{\rm vir},$ of SZ halos as a function of redshift, $z,$ for the ACT experiment compared to a compilation of groups from the USGC \cite{ramella_2002} (plotted as crosses) and a galaxy group catalog based on the Sloan Digital Sky Survey (SDSS DR4)\cite{yang_2007} (plotted as dots). The SDSS groups are only shown up to a redshift of $z=0.1$. The minimum detectable mass is plotted for an entropy model with $S/N=5$ (solid curve) and $S/N=3$ (dashed curve).}
  \label{halo_detection_limits_vs_data}
\end{figure}

In Fig.~\ref{halo_detection_limits_vs_data} we compare detection curves for the ACT experiment to the distribution of nearby groups in the USGC (UZC-SRSS2 Group Catalog) \cite{ramella_2002} and a group catalog compiled from the Sloan Digital Sky Survey (SDSS) \cite{yang_2007}. The USG catalog is based on the updated Zwicky Catalog (UZC) and Southern Sky Redshift Survey (SRSS2), and contains 1168 groups of galaxies out to a redshift of $z \simeq 0.04$ and over a solid angle of 4.69 sr, while the group catalog based on the SDSS contains 301237 groups from $z=0.02$ to $z=0.2$. Comparing the ACT detection curves to the USGC we find that there are 429 groups and clusters in the catalog out to a redshift $z=0.04$ that can be detected at a level $S/N=5,$ with 222 of these halos having a mass below $2 \times 10^{14} M_\odot.$ At a signal to noise level of $S/N=3$ the numbers increase to 488 halos with 281 of these halos having a mass below $2 \times 10^{14} M_\odot.$ At redshifts beyond those probed by the USGC the SDSS catalog contains additional groups that are above the minimum detectable mass level of SZ experiments. The comparison of the threshold mass curves to these catalogs suggests that there are a large number of galaxy groups already detected in redshift surveys that can be detected through targeted observations with upcoming SZ experiments. 

\begin{table}[t!]
	
	\begin{tabular}{|c|c|c|c|} \hline
	Experiment & Model & $S/N=3$  &  $S/N=5$  \\ \hline
	ACT Deep & Entropy & 1520 & 696 \\
	ACT Deep & Polytropic & 3570 & 813\\
	ACT Wide & Entropy & 307 & 23 \\ 
	ACT Wide & Polytropic & 1371 & 87 \\ \hline
	SPT Wide & Entropy & 5487 & 131 \\
	SPT Wide & Polytropic & 15856 & 759 \\
	SPT Deep & Entropy & 2323 & 706 \\ 
	SPT Deep & Polytropic & 3845 & 1465 \\ \hline
	PLANCK & Entropy & 653 & 227  \\
	PLANCK & Polytropic & 1110 &  418 \\ \hline
	\end{tabular}

\caption{Anticipated yield of SZ halos detectable by the ACT, PLANCK and SPT experiments, calculated
by integrating the cluster abundance from the minimum detectable mass for a given survey and $S/N$ level to an upper mass limit of $M_{\rm vir} = 2 \times 10^{14} M_\odot$.}
\label{table:yield}
\end{table}

\section{Discussion and Conclusions}

We have investigated the detectability of SZ groups using an analytic prescription for the hot gas in these halos. 
The models that we studied were based on hot gas being in hydrostatic equilibrium with the dark matter halo, and described by a polytropic equation of state, or an equation of state modified to include an entropy injection term. We have found that the entropy models are distinguishable from the polytropic models via measurements of their SZ distortion, even in low mass clusters and galaxy groups. While these models provide a useful starting point to evaluate the detectability of SZ groups, an improved analysis will include a more realistic treatment of the gas distribution as provided by high resolution cosmological simulations, which we intend to pursue in a forthcoming paper. 

Another issue that we have only partially addressed here, through the inclusion of an SZ background contaminant, is the confusion caused by the superposition of SZ distortions from hot gas in halos along the line of sight (see for e.g., \cite{holder_2007, hallman_2007b}). Here again a large volume cosmological simulation will help to quantify the impact of the SZ background on the detection of group halos and their recovered flux. By taking advantage of the fact that nearby group halos produce a more extended SZ signal, we aim to mitigate the impact of the SZ background by devising algorithms to separate the group signal from the SZ emission at smaller angular scales produced by higher redshift clusters. Finally the combination of maps of the various foreground contaminants with simulated SZ maps will allow us to undertake a more accurate treatment of the foreground contamination. While we have been relatively conservative in our modelling of the foreground contaminants, we have not included effects such as the clustering of infrared point sources, which could turn out to be a significant contaminant in the extraction of SZ halos  \cite{righi_2008}. We have also not included the kinetic SZ effect as a possible contaminant because it has
the same frequency dependence as the primary CMB but a much smaller amplitude on the relevant angular scales. For the same reason we have not attempted to detect the halo via its kinetic SZ signal, though we note that a detection of the kinetic SZ signal could be enhanced by cross-correlation with optical or X-ray observations of the halo or the thermal SZ signal.

Prospects for detection of SZ clusters have been studied previously in the case of PLANCK \cite{schaefer_2006, melin_2006}, ACT \cite{pace_2008, sehgal_2007} and SPT \cite{melin_2006}, but these studies have mainly focused on the statistics of SZ detections above the mass completeness limit of the respective surveys. Pace et al. \cite{pace_2008} found that ACT could detect SZ halos down to $6\times 10^{13} h^{-1} M_\odot$ fairly independent of redshift, whereas we have found that these halos become hard to detect at higher redshifts. The analysis of \cite{pace_2008} only included the CMB as a foreground so it is conceivable that the inclusion of point source foregrounds would degrade their forecasts for low mass halos at high redshift, as we have found to be the case in our analysis. We also note that the analysis of \cite{schaefer_2006} found that PLANCK could detect halos of mass $6\times 10^{13} M_\odot/h$ below $z=0.1$ when they included the CMB and all galactic foregrounds. In our analysis we have emphasised that smaller halos, with $M \simeq 3-4\times 10^{13} M_\odot,$ could be detected at $z~\lsim~0.01$.

The detection of hot gas in these galaxy groups and low mass clusters with the upcoming set of SZ survey experiments will provide an interesting probe of galaxy formation and its effect on the distribution and state of the hot gas, which is most prominent in these halos. A measurement of the SZ distortion at the virial radius of these halos will set a joint constraint on the level of entropy injection and the baryon fraction, which can be compared to the predictions from galaxy formation models. In particular a measurement of the baryon fraction in the outskirts of galaxy groups and low mass clusters will provide a unique update to the baryon census in the local universe. 

While the upcoming generation of SZ experiments have been designed to carry out blind surveys of galaxy clusters, a targeted survey of galaxy groups already detected in optical and X-ray observations would yield a very interesting set of objects to study. The measurement of a diffuse signal on the scale of tens of arcminutes will be challenging though, and carefully planned and executed observations will be necessary to control systematic effects and produce high fidelity maps of the SZ distortion in these halos. In combination with existing optical and X-ray observations of these halos, these measurements will enhance our knowledge of the physics of galaxy formation and its effect on the intragroup medium.

\begin{figure}[!t]
\includegraphics[width=0.45\textwidth]{\plotsdir 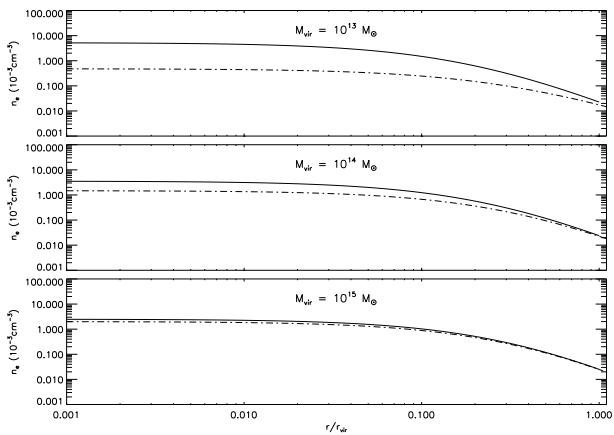}
\caption{Electron number density profiles, $n_e$ ($10^{-3}\mbox{cm}^{-3}$),  
plotted against $r/\Rvir$ for the polytropic model (solid curves) and entropy model
(dot-dashed curves). The panels from top to bottom show the density profiles 
for halo masses $M_{\rm vir}=10^{13},10^{14}$ and
$10^{15}~M_\odot$ respectively.} \label{fig:density_profiles}
\end{figure}

\begin{figure}[!t]
  \includegraphics[width=0.45\textwidth]{\plotsdir
    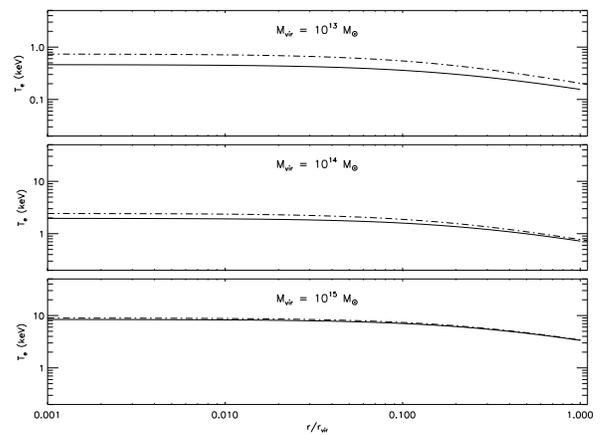}
  \caption{Electron temperature profiles, $T_e$ (keV), plotted against $r/\Rvir$ for
    the polytropic model (solid curves) and the entropy model (dot-dashed curves). 
    The panels from top to bottom show the temperature profiles for halo masses 
    $M_{\rm vir}=10^{13},10^{14}$ and $10^{15}~M_\odot$ respectively.}
  \label{fig:temperature_profiles}
\end{figure}

\begin{acknowledgments}
We thank Neelima Sehgal and David Spergel for useful discussions during the course of this project. 
KM and NG acknowledge financial support from the National Research Foundation (South Africa).
HT is supported by an Institute for Theory and Computation Fellowship.
RW acknowledges the SKA project office (South Africa) for financial support during his PhD.

 \end{acknowledgments}


\appendix*
\section{Gas profiles}

\begin{figure}[!t]
  \includegraphics[width=0.45\textwidth]{\plotsdir
    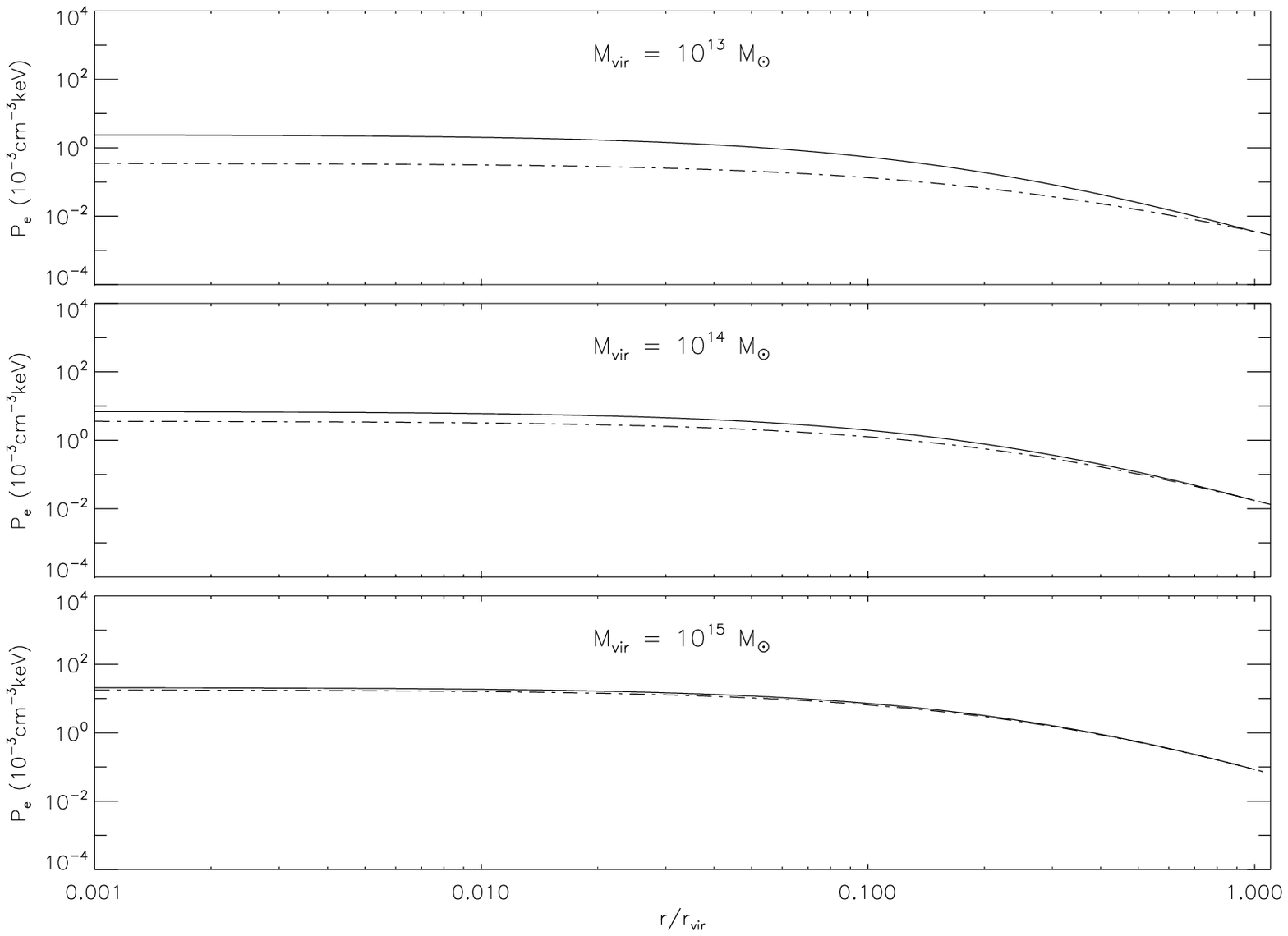}
  \caption{Electron pressure profiles, $P_e$
    ($10^{-3} \,\mbox{keV}\,\mbox{cm}^{-3}$), plotted against $r/\Rvir$ for the
    polytropic model (solid curves) and the entropy model (dot-dashed curves). The panels 
    from top to bottom show the pressure profiles for halo masses 
    $M_{\rm vir}=10^{13},10^{14}$ and $10^{15}~M_\odot$ respectively.} 
	\label{fig:pressure_profiles}
\end{figure}

\begin{figure}[!t]
  \includegraphics[width=0.45\textwidth]{\plotsdir 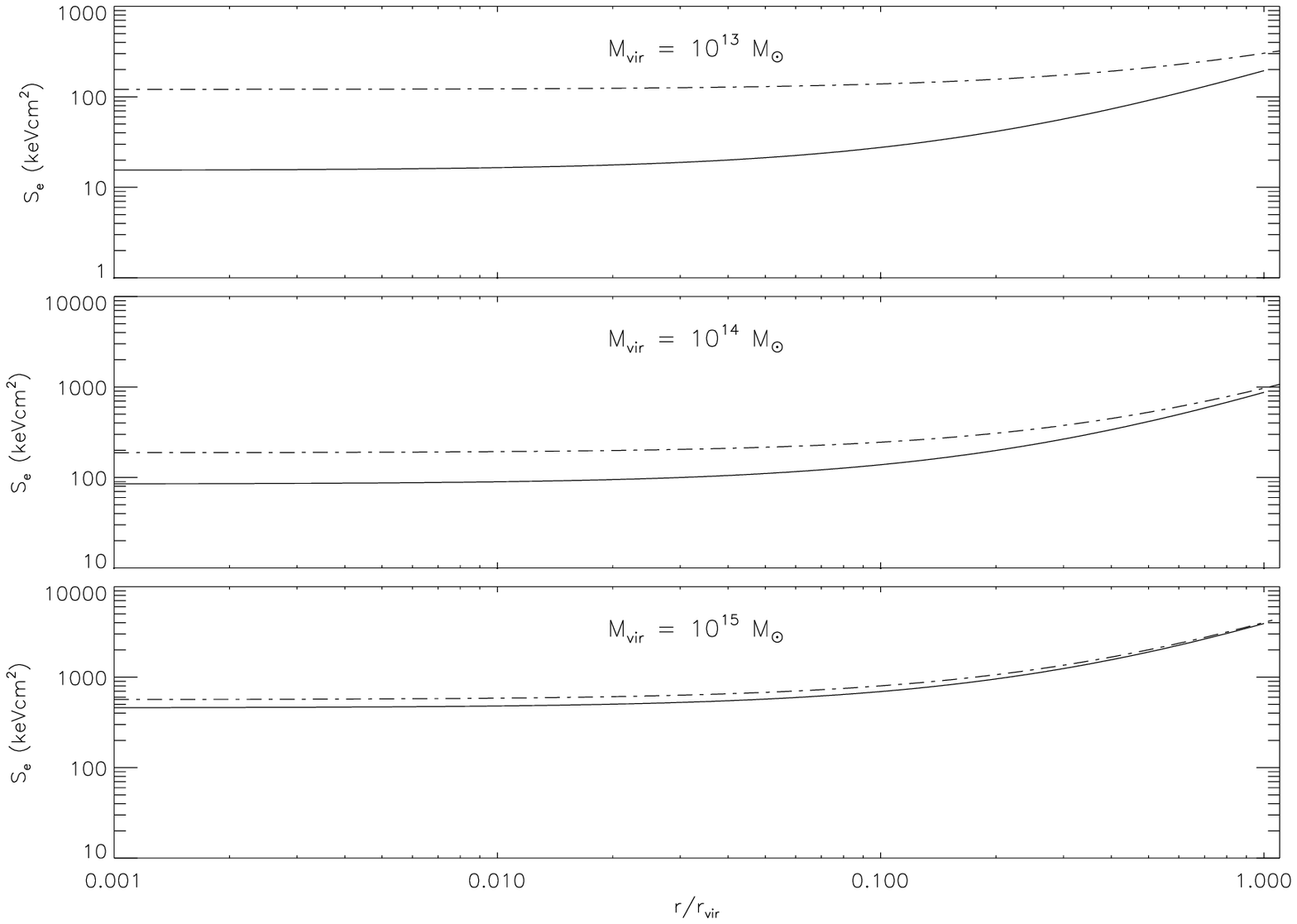}
  \caption{Entropy profiles, $S_e$ ($\mbox{keV}\,\mbox{cm}^2$), plotted against
    $r/\Rvir$ for the polytropic model (solid curves) and the entropy model
    (dot-dashed curves). The panels from top to bottom show the entropy profiles for halo masses 
    $M_{\rm vir}=10^{13},10^{14}$ and $10^{15}~M_\odot$ respectively.}
  \label{fig:entropy_profiles}
\end{figure}

We present here the radial profiles of the electron temperature, density, pressure
and entropy for the polytropic model and entropy injection
model (see Figs.~\ref{fig:temperature_profiles}-\ref{fig:entropy_profiles}).  
For the higher mass halos ($ \sim
10^{15}M_\odot$), the entropy injection model profiles are
very similar to the corresponding polytropic model profiles, which indicates
that the distribution of gas in large clusters is fairly insensitive to
the injection of entropy.

\begin{figure}[!t]
  \includegraphics[width=0.45\textwidth]{\plotsdir
    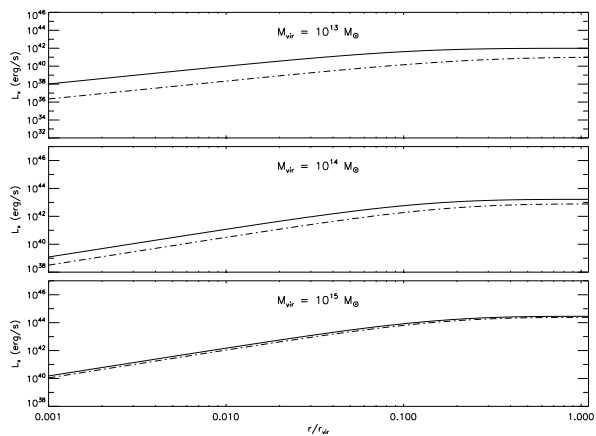}
  \caption{Integrated X-ray luminosity, $L_X (< r/\Rvir)$, plotted against $r/\Rvir$ for the polytropic model  
 	(solid curves) and the entropy model
    (dot-dashed curves). The panels from top to bottom show the integrated luminosity profiles for halo masses 
    $M_{\rm vir}=10^{13},10^{14}$ and $10^{15}~M_\odot$ respectively.}
  \label{fig:x_ray_profiles}
\end{figure}

For lower mass halos, the imposed heating is much more effective,
resulting in a higher electron temperature especially in the central
parts of the halo (see Fig.~\ref{fig:temperature_profiles}). This
reflects the fact that feedback effects are more significant
for galaxy and group sized halos, raising the temperature above the shock-heated
infall value.  The density profiles of the polytropic and entropy
injection models are significantly different for the low halo mass
range, demonstrating that the entropy injection has a more marked effect 
on group sized halos, pushing gas into the outer regions
of the halo and flattening the density profile. There is much more
hot gas in the inner halo regions in the polytropic model which
produces levels of X-ray emission in galaxy sized halos that are in
violation of observational constraints, as we discuss below.
Similarly, the imposed heating only significantly alters the pressure
profiles for galaxy and group sized halos (see Fig.~\ref{fig:pressure_profiles}),
greatly lowering the central electron
pressure. The heating term was modelled such that the pressure was
unaltered at the virial radius, where we do not expect feedback to have
an effect even for the lowest mass halos.  The entropy injection model
was constructed to have significantly more entropy than the
polytropic model in the inner regions for the low mass halos, where
the injected energy input is significant relative to the gravitational
binding energy of the halo.

The cumulative X-ray luminosity profiles for the two models are shown
in Fig.~\ref{fig:x_ray_profiles}. While the integrated luminosity
profiles are similar for high mass halos, the X-ray luminosity in the
polytropic model is significantly higher for low mass halos, in
violation of the upper limits on the diffuse X-ray emission from hot
gas in nearby galaxy halos like M31 \cite{taylor_2003,Takahashi_2001}. 
The entropy model does not violate these
constraints though, as the reduced central density in this model
lowers the X-ray luminosity in the inner regions despite
the increased temperature.


\bibliography{sz_paper}

\end{document}